# We just cannot have classical and quantum behavior at the same TIME

Knud Thomsen, Paul Scherrer Institut, Switzerland, knud.thomsen@psi.ch

**Abstract**
*Frauchiger and Renner recently cast doubt on the universal applicability of Quantum Mechanics [1]. In the following, it is pointed out that their conclusion of one of three common-sense conditions, demanded for Quantum Mechanics, being inevitably and permanently violated is not inevitable; there is a way out. Here, it is argued for fully accepting encompassing complementarity and for a basic conceptualization of quantum mechanics, in which different aspects are exhibited under different circumstances. The reported result is taken as an indication of a fundamental arrow of time pointing in the direction from the quantum --> classical domain. With the origin of thermodynamic time at that interface, its unrestricted applicability in isolated quantum systems is questioned. The bold proposal is that TIME as we know it just does not exist at the quantum level. Time instead commences together with causality at all interfaces to the classic domain whenever a minimum amount of energy is transferred in an uncontrolled manner, i.e., without well-balanced compensation. Insisting on fully comprehensive self-consistency opens new perspectives on (retro-)causality and also on an inflation phase at the beginnings of our universe.*

In a recent paper attracting considerable attention, Daniela Frauchiger and Renato Renner describe a thought experiment extending a line of argument originally conceived by Eugene Wigner [1]. They arrive at a contradiction between the measurements for seemingly the same state obtained by two actors when strictly following a protocol of uncritically applying quantum mechanics, QM. This contradiction is presented as excluding one of these generally assumed (expected / hoped for) characteristics of quantum theory:

- **U**niversality, i.e., quantum theory is applicable at all scales and for all types of objects
- **C**onsistency, i.e., quantum theory allows independent agents to make predictions, which do not contradict each other
- **U**niqueness, i.e., opposite established facts cannot both be true

The proposed interpretation of this result by the authors and also by several other experts was that one of the above three seemingly innocent and very intuitive common-sense assumptions, all in essence valid in the classical realm (i.e., excluding quantum and relativistic effects), has to be abandoned [1,2].
Indeed, even without directly questioning the purported contradiction, there is an immediate fourth way out:

- **C**omplementarity, i.e., these three features simply cannot be observed at the same time; they apply in diverse conditions / under different perspectives, just not simultaneously in the set-up as described together and lasting for an extended period of time.

Frauchiger and Renner assert in their work how the main existing different proposals of interpreting quantum mechanics fare with respect to the three conditions of Universality, Consistency and Uniqueness. They conclude that each serious attempt foregoes one of "UCU".

In the following, it shall first be argued for "UCUC", i.e., that invigorating a fundamental concept of Complementarity as devised in the earliest attempts at understanding queer quantum phenomena can also serve at this effective meta-level to bring some order in an apparently contradictory overall frame.
In the light of the many fine-grained approaches, which have already been proposed, it is deemed necessary to first take a big step back and lay out a mosaic, even if vaguely and with gaps, to coarsely chart the territory before later delving into exact details. A picture is sketched in very broad strokes admitting that the findings of Frauchiger and Renner actually are to be expected in the light of the many-faceted enigmas of quantum mechanics.
As with combining particle- and wave- characteristics, with a compound conceptualization of diverse, seemingly contradictory aspects, one can hope for putting together an overall puzzle by using all observable pieces in their appropriate place and at the correct time.

As a starting point, complementarity is taken as meaning that particle- and wave- aspects of physical objects cannot be observed fully at the same time. The proposed extension is that the three common-sense conditions (UCU) as described in the Gedankenexperiment cannot concurrently be met in one time-independent context.

Frauchiger and Renner in their interpretation tend to discard the universality of quantum mechanics as no clear-cut experimental evidence for its applicability seems available for large masses.
It can be argued that, e.g., the Chandrasekhar limit for the allowed mass of white dwarfs is a convincing case of some successful application of QM at the scale of several solar masses in a quite chaotic state, which can be relatively directly observed without any disturbance to the objects and using rather simple classical means [3]. Therefore, it is suggested to put weight rather on complexity, particularly on some type of comprehensive self-reference, and intimately associated with that, time [4].

The authors' proposal to possibly turn their thought experiment in the future into a real one employing quantum computers, actually refers to a fundamentally different constellation than the one of the Gedankenexperiment as it involves the full reversibility of computers acting as observers.
This latter condition of taking such reversibility for granted, here is seen as the crux of the thought experiment and the deep-lying origin of the derived contradictions.

Embracing complementarity accepts that any real object, which behaves according to classical mechanics, CM, will not just arbitrarily switch character or simply be made to exhibit full quantum behavior. This asymmetry can be conceptualized as resulting from different versions of decoherence and/or some form of collapse of a wave function but there is no immediate need to formally specify any exact mechanism. For a very first step it suffices to simply acknowledge the two domains of classical and quantum behavior as distinct and the transition between them as being asymmetrical.

This presupposes as well as implies an arrow of time, i.e., a default direction from QM --> CM, and it actually forbids the full performance of the suggested measurement protocol; in particular, putting a classically described entire laboratory including an experimenter in a superposition state or reversing their history just poses a practically insurmountable problem. We simply do not have sufficient control over all necessary microscopic details and cannot self-consistently expect to ever achieve that.

The mere existence of two well-distinguished but intimately intertwined domains, one fragile and the other robust, as actually observed, is sufficient to establish a preferred direction for changes, i.e., for starting the flow of time.

Nicely, the so founded arrow of time generally is aligned with all the others surfacing in areas like cosmology, thermodynamics and psychology, and it also matches with difficulties describing joint states of a composite of systems at different times, as well as an inherent difference in the treatment of space and time inside QM [5]. What has been presented as arrow of time of familiar quantum mechanics amongst others by Murray Gell-Mann and James Hartle, in fact assumes a fixed background (Newtonian) spacetime to start with (extrapolated from CM) and thus is not necessarily intrinsic to QM [6].

So, the proposal here is to first accept the result of Frauchiger and Renner at face value for the sake of the argument and to take it as an indication of a fundamental arrow of time. The phase transition between behavior following the rules of QM to one obeying classical mechanics evolves naturally in the QM --> CM direction, and the reverse, if at all, can only be enforced with great effort. Quantum states, of which we know with some certainty, need to be carefully prepared and isolated from any disturbing environment.

In addition to their fragility, given the fundamental fuzziness and uncertainty exhibited by quantum mechanical entities, it is no wonder, that this transition interface is fuzzy as well; some type of time-energy uncertainty relation undoubtedly applies in general.

The passage of time then is witnessed and, due to overwhelming statistical probabilities, irreversibly recorded by the trajectory / memory of successively established classical events and facts. Trying to apply the wrong description at one or the other side of that divide, self-consistently results in paradoxes and contradictions. The consistency of classical observations obviously reaches back to the QM world but only to some extent. The Schrödinger equation is valid for conservative systems and its basic version is time-independent. The time dependent Schrödinger equation then involves a first order derivative in time and thus defines some time-ordering, but this is not the case between superpositions of states of an isolated system, and it is not visible in detail to the CM outside (Born's rule can be seen a special procedure for some type of time averaging of ergodic processes [7]), and there is no scale with globally defined units, in particular, not including superpositions.

Not to speak of crossing back and forth, the mere fixing of the transition-point between CM and QM has been described as tricky before [8]. Some of this has been discussed time ago under the name of "shifting split". This fits nicely with comments on Frauchiger and Renner by Franck Laloë and Anthony Sudbery, who both point out that in order to arrive at the reported no-go result no consistent application of QM rules, and, in particular, no consistent specific time-points for the measurements are applied in their Gedankenexperiment [8,9].

Basically the same conclusion of an inconsistent application of the collapse rule has been reached by Dustin Lazarovici and Mario Hubert [10].

A most recent state-of-the-art 6-photon experiment purportedly rejecting observer-independence in the quantum world suffers from the very same collapsing of actions and events, which (crucially) are advertised as happening one after the other in classical time, into effectively a single point in time [11].

The various extant interpretations of QM in this view are all valid, and also, –not(!) at the same time–, (partially) invalid; it strictly depends on the constellation and the full context, which perspective applies best. To be accepted as principally valid candidates, the underlying formalisms need to yield the same observable results irrespective of the particular interpretation; they need to be equivalent from a well-defined coarse observational perspective.

This might only at first sight be taken as argument for QBism or relational quantum mechanics, which plainly dismiss overall consistency [12,13]. The argument here, in fact on a meta-level, goes significantly beyond an assertion that any measurement would only be real for the involved agent; it rather suggests a conception where causal structure matters and even a high-level union of kinematics and dynamics strictly inside QM is of limited relevance [14]. If any of the well-known interpretations of QM seem to show promise for better fit with the considerations here, these are accounts along the consistent / decoherent history approaches (added a comprehensively self-consistent sense of time) and similarly augmented Bohmian mechanics [6,10,15,16].

The decisive step proposed here thus is to seriously reflect the entire universe (including (in case) actors and observers), which is solidly grounded in observation, back on itself and to stress full encompassing self-consistency in the experienced unique real world. This is the very best obtainable from inside the accessible universe anyway. Naive unbiased observation tells that there is one universe, –widely interacting in our causally connected region in space over distances vastly exceeding galactic dimensions–, and that, to a very large extent, it has behaved since about 13.8 billion years and still behaves on the most relevant scales as well-described by Quantum Mechanics, Classical Physics and Special as well as General Relativity; and this one world in practice is characterized by strong time asymmetries.

The macroscopic universe (its expansion) and the structures in it, which have evolved over time, definitively bear witness to classical physics and (thermodynamic) time irreversibility long before the emergence of any conscious observers and perfectly doing without any. Some type of decoherence in a structured environment with asymmetric interactions and / or boundary conditions obviously is good enough. Once first classical facts have manifestly been established, by sheer statistics and involving exponentially unbalanced probability ratios the impossibility of running some coarse version of universal (classical) time backwards ensues, and these facts can serve to support the further development in one well-defined coarse grained history without getting lost in endless branching, not even in many parallel and incompatible narratives. A major ingredient in the ETH interpretation, some principal loss of access to the past somewhat akin to decoherence or diminishing potentialities, needs not to be postulated as additional principle [17], it can modestly be understood as a fact observed in nature and consequence of the asymmetric relation between QM and CM. Except in very dedicated and sophisticated experiments, temporal support points for the grand history are not assigned, but incessantly result during the course of the universe's evolution over classical time.

There is no absolute time in Einstein's special relativity, simultaneity is relative. In accordance with the applicable reference frames, the experienced order of causally disconnected events (only) can vary for different observers. Ubiquitous identical seeding for the origin of time ensures that it runs in the same direction everywhere and makes the proposal here dovetail nicely with Special Relativity.

Assuming the first origin of time at the very transitions from QM --> CM matches perfectly with the observed arrow of time in our (the classical experimenters') real commonsense world while it implies that this concept might not be fully applicable in the quantum domain with objects most often best described as being in a superposition of entangled states.
The bold proposal then is that TIME as we know it just does not exist for isolated quantum systems. With "timelessness" at the quantum level, all possible states exist together. Only when sufficiently disturbing a QM system, clear-cut "measurement results" become visible to and in the outside classical world. God does not throw the dice; it is rather that the environment, in particular, in measurements: we draw tickets from the full basket of the rich lottery of possibilities.
As a condition for crossing the QM --> CM boundary a minimum amount of energy transfer is suspected. It can be assumed, that intricate conditions apply with respect to the relevant short time frames and to what extent the interaction is controlled, i.e., disturbances can possibly be compensated by involving additional degrees of freedom. Some inspiration for how this might be accomplished and described in detail can certainly be drawn from the results of Rolf Landauer on the cost of computing [18]. A first simple example by Bertúlio de Lima Bernardo delivers limits for short and long wavelengths corresponding to high or low energy transfer in time correlation decoherence [19].

The asymmetry in the transition probability QM --> CM and associated minimum durations avoid problems with any possibly purported circularity, – of reality as well as of the argument. Observations often have an effect on the measured value but this can, if at all, only fully be observed at a later point in classical time. The widest achievable and ever increasing consistency of measurements and predictions is the hallmark of advancing science at the most detailed level as well as overall; this includes elucidating conditions in which meaningful predictions just are not possible.

Causality as we know breaks down in the absence of a time ordering; our classical preconceptions and commonsense habits are not applicable inside the pure quantum world.
Retrocausality has been found to be a mandatory ingredient in realistic time symmetric interpretations of QM [20]. Whereas "no (classical) time passing" is a special symmetry condition, "timelessness" might rescue some minimum version of compound realism, and save uniqueness and free choice (from accessible options) without implying disturbing retrocausality (which would be effective in the real world).

Decoherence has been proposed to be responsible for an arrow of time effectively out of the quantum realm (while formally staying inside); it is observed from / on the CM outside [21]. The same is true for the sudden death of entanglement including its rebirth, and the difference between these two phenomena might be taken as more evidence that classical time just cannot be simply assumed as unaltered valid strictly inside QM [22]. The quantum realm features richer causal relations than CM; under the influence of local environmental noise, the quantum to classical transition for causal pathways is different depending on the mixing of diverse pathways, and the coherence in a mixture can be more sensitive than the coherence in the individual causal pathways [23].

An isolated externally non-interacting quantum system without internal clock would have no way of knowing that it has moved in an outside frame of reference. Spooky action at a distance does not exist for the internal perspective. Delayed choice experiments likewise enforce classical behavior as long as a choice is effected while the system is in limbo from the outside perspective. With a one-way act of measuring (external, CM) even with (internally QM) no time passing, no signal is ever sent to the past.

Keeping on that coarse level and taking a generic uncertainty relation $\Delta E * \Delta t \geq \hbar$; i.e., in the absence of energy transfer no time passes ($\Delta E = 0 \wedge \Delta t = $ "$\infty$").

Likewise, cooling to very low temperatures and isolating a system would provoke and maintain quantum behavior pushing uncontrolled energy exchange to below a maximum threshold.

Time is not a trivial observable in QM and no time-energy uncertainty relation is universally valid [5,24,25]; this does not come as a surprise assuming that our usual concept of time simply cannot be applied there.

High-precision experiments with quantum systems, which apparently carry some type of (expectedly hidden) internal clocks, e.g., delayed-choice or double-slit experiments employing muons, some of which decay during their flight at different positions / at different external times, might be interesting to perform. An obvious prediction is that the decay interrupts the system like an external intervention and this can be seen in the distributions of the resulting electrons and photons. No high-energy experiment could be analyzed without the ability of tracing detected particles rather precisely back to their respective origins. For decaying particles, which move with velocities close to the speed of light, time dilation as described in Special Relativity extend the observed life times.

At a singularity ("Big Bang"), no fixed spacetime geometry or time order can be assumed, and the absence of a pure state might be the right starting condition for later decoherence [6]. Concerning the very beginning of time, it is thus tempting to speculate that some type of inflation period is only marked at its end when a threshold of separation / distinction on the way from QM to CM is exceeded, with no mandatory superluminal expansion phase but timelessness before classical time sets in. So, one should not prematurely try to amend Wheeler-DeWitt with a time parameter. In a naive continuous picture, time dilation due to (almost) infinitively strong gravity and then during expansion with the speed of light could match consistently with timelessness in the quantum realm at the beginning of the universe. From this point of view no vast expansion over a very short time interval resulting in widely separated identical features on a smooth sky has to be explained, it is rather that in a homogenous soup with little effective energy transfer not much classical time passed before reheating. Special initial conditions and the asymmetry QM --> CM combined in the seeding of time; a final global effective standstill maybe (no time passing in a cold and widely diluted universe), but no full reversal of its direction in the future appear to be plausible scenarios.

An effective absence of time in conditions where the common-sense concept of time does not apply contributes a profound reason why it turns out to be difficult bringing quantum mechanics and General Relativity together in one common comprehensive edifice.
In the wide space of the expanded universe, gravity could in the end even be a fundamental cause for further directing the advancement of time by incessantly pulling and nudging systems from the quantum to the classical realm [26]. Searching for gravitational effects in fully isolated quantum systems might be self-contradictory, and it might also be speculated that no time can be dilated when there is none.
It is not necessary but neither disturbing, that the arrow of time also emerges naturally for an internal observer in an uncomplex system of N point particles interacting through Newtonian gravity [27].

As a first step, admittedly, the very coarse overarching frame as sketched here does not just solve once and for good all the issues associated with the various interpretations of quantum theory discussed for the last century; neither does the complementarity of wave- and particle-descriptions for the comparatively simple case of a single particle. Still, leaving solipsistic points of view aside, there exists one real observable world, which physicists strive to describe and comprehend. As no true outside perspective on the universe is rationally possible, we'd rather aim for encompassing self-consistency. Accepting the established queerness of QM for full and understanding our real experienced world as resulting from a truly fundamental asymmetry in the transitions between the domains of QM and CM, as well as working out – if not close – some gaps and interfaces, might establish some gain. This endeavor including the sharpening of differences between distinct interpretations could allow for some progress in the form of an overall reconciliation of quite diverse extant approaches, putting them into perspective and make them shed their light on these topics from unequal points of view. Such a common comprehensive conceptual basis, effectively on a meta-level, hopefully turns out useful and best suited for approximations with their respective limits of applicability and aiming at different purposes under significantly different conditions.

Returning to the questions raised a century ago and newly tackled by Frauchiger and Renner, the main answer proposed here can in short be paraphrased as "U̶CUC"; quantum mechanics is not universally applicable. There is a distinct classical world with time flowing only there. In the quantum realm, there is no intrinsic time; it commences its flow solely at the very interface to our classical real world at each occasion when an uncontrolled disturbance occurs, and effectively some minimum amount of energy is irreversibly transferred.

## Acknowledgments

Diverse very thoughtful and encouraging comments by a number of interested experts and reviewers are gratefully acknowledged. Rather than trying to substantially rewrite the current sketch, it shall be attempted to present some of its essence in a new and clearer paper presenting more stringent and quantitative, while still comprehensive, arguments.

## References

1. D. Frauchiger and R. Renner, Quantum theory cannot consistently describe the use of itself, Nature Communications volume 9, Article number: 3711, 2018. https://doi. org/10.1038/s41467-018-05739-8

2. A. Ananthaswamy, New Quantum Paradox Clarifies Where Our Views of Reality Go Wrong. https://www.quantamagazine.org/frauchiger-renner-paradox-clarifies-where-our-views-of-reality-go-wrong-20181203/

3. S. Chandrasekhar, The Maximum Mass of Ideal White Dwarfs, Astrophysical Journal 74, 81-82, 1931.

4. M.L. Dalla Chiara, Logical self-reference, set theoretical paradoxes and the measurement problem in quantum mechanics. J. Philos. Log. 6, 331-347, 1977.

5. D. Horsman, C. Heunen, M.F. Pusey, J. Barrett, and R.W. Spekkens, Can a quantum state over time resemble a quantum state at a single time? Proc. R. Soc. A 473, 20170395, 2017.

6. M. Gell-Mann and J.B. Hartle, Time Symmetry and Asymmetry in Quantum Mechanics and Quantum Cosmology, arXiv:gr-qc/9304023v2, 21 Jun, 2005.

7. A. Khrennikov, Born's rule from measurements of classical signals by threshold detectors which are properly calibrated, Journal of Modern Optics, 59:7, 667-678, 2012, DOI: 10.1080/09500340.2012.656718 .

8. F. Laloë, Can quantum mechanics be considered consistent? a discussion of Frauchiger and Renner's argument, arXiv:1802.06396v3, 9 July, 2018.

9. A. Sudbery, Single-World Theory of the Extended Wigner's Friend Experiment, Foundations of Physics 47, 658-669, 2017.

10. D. Lazarovici and M. Huber, How Quantum Mechanics can consistently describe the use of itself, Nature, 24 Jan. 2019, DOI:10.1038/s41598-018-37535-1.

11. M. Proietti, A. Pickston, F. Graffitti, P. Barrow, D. Kundys, C. Branciard, M. Ringbauer, and A. Fedrizzi, Experimental rejection of observer-independence in the quantum world, arXiv:1902.05080v1, 13 Feb 2019.

12. C. Rovelli: Relational Quantum Mechanics, Int. J. Theor, Phys. 38, 1637-1678, 1996.

13. C.A. Fuchs, D.N. Mermin, and R. Schack, An Introduction to QBism with an application to the locality of quantum mechanics, A. J. Phys. 82, 749-754, 2014.

14. R.W. Spekkens, The paradigm of kinematics and dynamics must yield to causal structure, in A. Aguirre et al. (eds.), Questioning the Foundations of Physics, The Frontiers Collection, Springer, 2015, DOI 10.1007/978-3-319-13045-3_2.

15. R.B. Griffiths, Consistent histories and the interpretation of quantum mechanics, J. Stat. Phys. 36, 219-272, 1984.

16. R. Omnès, Consistent interpretations of quantum mechanics, Rev. Mod. Phys. 64, 339-382, 1992.

17. P. Blanchard, J. Fröhlich, and B. Schubnel, A "garden of forking paths" – The quantum mechanics of histories of events, Nuclear Physics B 912, 463-484, 2016.

18. R. Landauer, R. Irreversibility and heat generation in the computing process, IBM J. Res. Dev. 5, 183-191, 1961.

19. B. de Lima Bernardo, Time correlation and decoherence in a two-particle interferometer, Brazilian Journal of Physics 44 (2-3), 202-207, 2014.

20. M.S. Leifer and M.F. Pusey, Is a time symmetric interpretation of quantum theory possible without retrocausality? Proc. R. Soc. A 473, 20160607.

21. H.D. Zeh, On the Interpretation of Measurement in Quantum Theory, Foundations of Physics, vol. 1, 69-76, 1970.

22. Ting Yu, J H. Eberly, Sudden Death of Entanglement, Science 323, 598-601, 2009.

23. K. Ried, J-P.W. MacLean, R.W. Spekkens, and K.J. Resch, Quantum to classical transitions in causal relations, arXiv:1707.06131v1, 19 Jul, 2017.

24. P. Bush, The Time-Energy Uncertainty Relation, Lect. Notes. Phys. 734, 73-105, 2008.

25. K. Urbanowski, Critical look at the time-energy uncertainty relations, 2019, arXiv:1908.05273v1

26. G. Gasbarri, M. Toroš, S. Donadi, and A. Bassi, Gravity induced wave function collapse, Phys. Rev. D 96, 104013, 2017.

27. J. Barbour, T. Koslowski, and F. Mercati, Identification of a Gravitational Arrow of Time, Phys. Rev. Lett. 113, 181101, 2014.